\documentclass[3p,times,twocolumn]{elsarticle}
\usepackage{amssymb}
\usepackage{graphics}
\usepackage{psfrag}
\usepackage{epsfig}
\usepackage{array,multicol,url}

\usepackage[dvipsnames]{xcolor}

\begin{document}

\begin{frontmatter}
\title{The proton magnetic radius: a new puzzle?}

\author[A]{Yong-Hui Lin}
\author[B,C]{Hans-Werner Hammer}
\author[A,D,E]{Ulf-G. Mei{\ss}ner}

\address[A]{Helmholtz Institut f\"ur Strahlen- und Kernphysik and Bethe Center
   for Theoretical Physics, Universit\"at Bonn, D-53115 Bonn, Germany}
\address[B]{Technische Universit\"at
   Darmstadt, Department of Physics, Institut f\"ur Kernphysik, 64289 Darmstadt, Germany}
\address[C]{ExtreMe Matter Institute EMMI and Helmholtz Forschungsakademie
   Hessen f\"ur FAIR, GSI Helmholtzzentrum f\"ur
   Schwerionenforschung GmbH, 64291 Darmstadt, Germany}
\address[D]{Institute for Advanced Simulation and Institut f{\"u}r Kernphysik,
            Forschungszentrum J{\"u}lich, D-52425 J{\"u}lich, Germany}
\address[E]{Tbilisi State University, 0186 Tbilisi, Georgia}

\begin{abstract}
We comment on the puzzling status of the proton magnetic radius determinations.
\end{abstract}  

\end{frontmatter}

The electric radius, $r_E$, and the magnetic radius, $r_M$, of the proton
are fundamental quantities of
low-energy QCD, as they are a measure of the probe-dependent size of the
proton.
Both radii are defined by the slope of the respective proton form factors
$G_E (t)$ and $G_M (t)$  at zero momentum transfer,
\begin{equation}
  r_{E/M}^2 = -\frac{6}{G_{E/M}(0)} \frac{dG_{E/M}(Q^2)}{dQ^2}\Big|_{Q^2 =0}~,
\end{equation}
with $Q^2=-t>0$ the invariant four-momentum transfer squared (defined to
be positive for space-like momentum transfer).
The form factors $G_E (t)$ and $G_M (t)$ describe the structure of
the proton as seen by an electromagnetic probe.
Following the groundbreaking work of Hofstadter and collaborators,
the form factors and radii have been measured in elastic electron
scattering experiments since the 1950's. A second avenue to measure the proton
structure is provided by measurements of the Lamb shift in hydrogen atoms,
which is also sensitive to the proton charge radius.
Lamb shift experiments in ordinary and muonic hydrogen have
lead to very precise determinations of the proton charge radius in
the last 30 years~\cite{Antognini:2013txn}.

While the electric radius of the proton has attracted much attention
in the last decade due to the so-called  ``proton radius puzzle'' (see, e.g.,
Refs.~\cite{Hammer:2019uab,Gao:2021sml} for recent reviews)\footnote{It is worth emphasizing that within the framework of dispersion relations, electron-proton scattering data always led to the so-called small radius of about $r_E \simeq 0.84$~fm \cite{Lin:2021umz}, consistent with the Lamb shift in muonic hydrogen.}, this is not true for its
magnetic counterpart. The magnetic radius is not probed directly in the
Lamb shift in electronic or muonic hydrogen and thus all existing information
comes from electron scattering experiments.  Note, however, that
the $ep$ cross section is dominated by the electric form factor
for small momentum transfer.
Thus the magnetic radius $r_M$ is more sensitive to larger momentum transfers,
and it is not known experimentally with the same precision as $r_E$.
Within dispersion relations, the value extracted for
$r_M$ was always bigger than 0.83~fm and slightly larger than $r_E$, with the most recent
high-precision analysis including data from the time-like region giving
$r_M = 0.849^{+0.003}_{-0.003}{}^{+0.001}_{-0.004}$~fm~\cite{Lin:2021xrc}
where the first error is statistical (based on the bootstrap procedure)
and the second one is systematic (based on the variations in
the spectral functions).
This value is in stark contrast
to the analysis of the A1 collaboration \cite{Bernauer:2010wm}, that led to the much
smaller value of $r_M = 0.777(13)_{\rm stat.}(9)_{\rm syst.}(5)_{\rm model}(2)_{\rm group}$~fm, including in addition to
statistical and systematic uncertainties also some uncertainties from the
fit model and differences between the two model groups used in the analysis.
However, looking at the
corresponding magnetic form factor $G_M(Q^2)$, it shows a pronounced bump-dip structure for
momentum transfers $0\leq Q^2 \leq 0.3$~GeV$^2$. Such a structure is at odds with
unitarity and analyticity \cite{Belushkin:2006qa}.

So is there other information available that could help to clarify this issue?
Indeed, lattice QCD calculations at physical pion masses are available. Let us
first concentrate on the isovector part of $r_M$. This is free of the difficult to compute
disconnected contributions and the pertinent results are given in Tab.~\ref{tab:rMiv},
together with the corresponding isovector electric radii.
\begin{table}  
\centering \begin{tabular}{|l|c|c|}
\hline
                   &  $r_E^V$ [fm]   &  $r_M^V$ [fm] \\
\hline
Disp. rel.         &  0.900(2)(2)    &   0.854(1)(3)  \\
\hline
PACS \cite{Tsuji:2023llh} &  0.832(19)(70)(22) & 0.771(64)(84)(10) \\
Mainz  \cite{Djukanovic:2023jag} &  0.882(12)(15)  &   0.814(7)(5) \\
Cyprus \cite{Alexandrou2022} &  0.920(19)(--)  &   0.742(27)(--) \\
Mainz \cite{Djukanovic:2021cgp}     &  0.894(14)(12)  &   0.813(18)(7) \\
PACS \cite{Shintani:2018ozy}     &  0.785(17)(21)  &   0.758(33)(286) \\
\hline
\end{tabular}
\caption{Electric and magnetic isovector radii from lattice QCD.\label{tab:rMiv}}
\vspace{-3mm}
\end{table}
We notice that there is at least a three-sigma deviation in $r_M^V$ between the dispersive
and the lattice results. This can also be seen in the value for $r_M$ given in
Ref.~\cite{Djukanovic:2023jag}, $r_M =  0.8111(89)$~fm, which roughly corresponds to a $4\sigma$
deviation from the dispersive value. Note, however,
that the electric radius in that work comes out rather small, $r_E = 0.820(14)$~fm. The recent
results by the PACS collaboration which include finite
lattice spacing effects \cite{Tsuji:2023llh}
also feature a small magnetic radius, but they 
have larger errors and are consistent with the dispersive value.

\begin{figure}
\begin{centering}
\includegraphics[width=0.95\columnwidth]{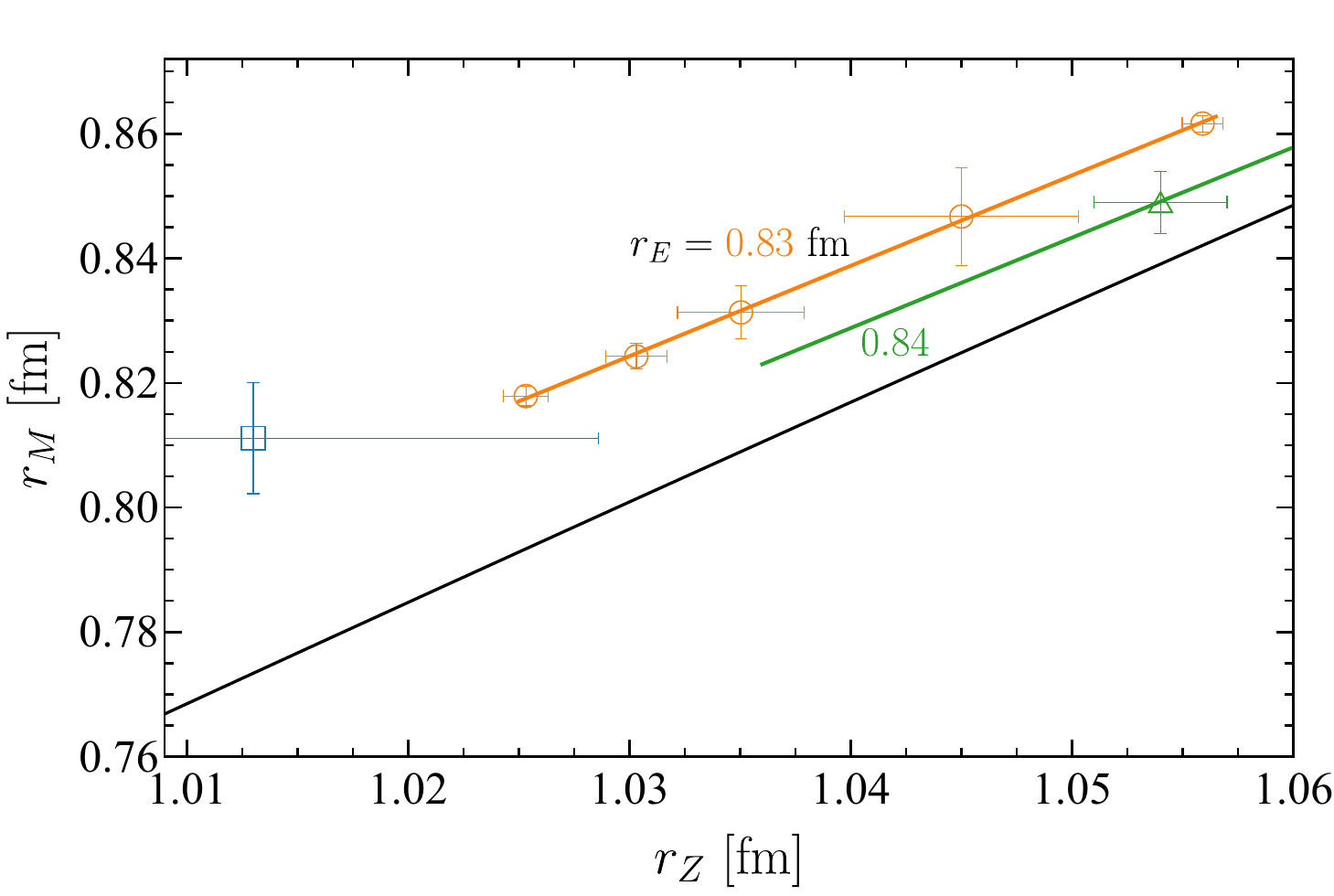}
\par\end{centering}
\caption{Dependence of the Zemach radius $r_Z$ on $r_M$. The black line shows the relationship 
	from the scenario of the generalized form factors with $r_E=0.83$~fm. 
	The open cycles show those by fitting to the PRad data
	and allowing $r_Z$ to vary in the range shown within the scenario of the dispersion theoretical form factors. The open square and triangle denote the latest LQCD determination by Refs.~\cite{Djukanovic:2023jag,Djukanovic:2023cqe}
	and the high-precision dispersive values~\cite{Lin:2021xrc}, respectively.
\label{fig:rZ}}
\vspace{-5mm}
\end{figure}

So how could one resolve this discrepancy? The most prominent way the
magnetic radius enters in atomic physics experiments is via the Zemach
radius $r_Z$, which enters in the Lamb shift. The Zemach radius is given by:
\begin{eqnarray}\label{eq:zmr}
  r_Z &=&-\frac{4}{\pi}\int_0^\infty \frac{\mathrm{d} Q}{Q^2} \left[\frac{G_{E}(Q^2)G_{M}(Q^2)}{1+\kappa}-1\right]~.
\end{eqnarray}
To explore the relation between $r_Z$ and $r_M$, we consider two types of the proton form factors,
the ones parameterized within the dispersion theoretical framework comprehensively introduced in 
Ref.~\cite{Lin:2021umz} and those using generalized dipole form factors,
that is
\begin{equation}
  G^D_{E,M}(Q^2) = \frac{G^D_{E,M}(0)}{(1+ Q^2/\Lambda_{E,M}^2)^2}\,,
\end{equation}
with $\Lambda_E \neq \Lambda_M$. On the one hand, using generalized dipole form factors allows for the
analytical calculation of the integral in Eq.~(\ref{eq:zmr}).
It reveals an approximate linear dependence on the magnetic radius within the range of 1.01~fm to
1.06~fm~\cite{Antognini:2022xoo}
by setting the charge radius to the PRad value of $0.83$~fm~\cite{Xiong:2019umf}, see the black line in
Fig.~\ref{fig:rZ}.
On the other hand, fitting to the PRad data and varying $r_Z$ with the dispersion theoretical parametrizations, 
a shifted linear relationship between the Zemach and the magnetic radius is indicated by the open cycles in
Fig.~\ref{fig:rZ}. 
Note that the charge radius is also fixed at the value of $0.83$~fm in these fits and the Zemach radius is constrained in the range of 1.025~fm to 1.055~fm by the PRad data. 
Furthermore, the linear relationship shifts from left to right when increasing the value of the fixed
charge radius to 0.84~fm, as also shown in Fig.~\ref{fig:rZ}.
The latest LQCD determinations from Mainz~\cite{Djukanovic:2023jag,Djukanovic:2023cqe} and
the high-precision dispersive values~\cite{Lin:2021xrc} 
are shown as the open square and triangle points, respectively. As can be seen quite clearly,
the precision of the current LQCD determinations of the proton form factors is still
not sufficient to resolve the puzzle.

We note that the high-precision dispersive analysis
\cite{Lin:2021xrc} gives $r_Z = 1.054^{+0.003}_{-0.002}{}^{+0.000}_{-0.001}$~fm,
consistent within $1\sigma$ with the value extracted from muonic hydrogen spectroscopy,
$r_Z = 1.082(37)$~fm~\cite{Antognini:2013txn}, but a more precise measurement would clearly be
very useful in further pinning down the proton magnetic radius. The recent lattice determination
of $r_Z$ \cite{Djukanovic:2023cqe} shows a clear deviation from the experimental value, which
can be traced back to their small proton electric and magnetic radii.

To summarize, we have shown that there are a number of conflicting determinations of the
proton magnetic radius, which could be named as ``the new proton radius puzzle''. In fact,
considering that the proton electric radius is known now, a very precise determination of the
Zemach radius would give another independent determination of $r_M$. Furthermore, in the
dispersive framework the Zemach radius comes out on the large end of the
range consistent with the muonic hydrogen and the  PRad data
and this leads to a large magnetic radius $r_M = 0.85$~fm.

\newpage

\vskip 0.3cm 
\noindent
{\bf Conflict of interest}
    
The authors declare that they have no conflict of interest.
\vskip 0.3cm 
\noindent
{\bf Acknowledgments}

We thank Franziska Hagelstein for a useful communication.
This work was supported in part by the Deutsche Forschungsgemeinschaft
(DFG, German Research Foundation) -- Pro\-jekt\-num\-mer 279384907 --
CRC 1245 and the DFG and the NSFC through
funds provided to the Sino-German CRC 110 ``Symmetries and
the Emergence of Structure in QCD'' (DFG project-ID 196253076 - TRR 110,
NSFC grant No. 12070131001). The work of UGM was also 
supported by the Chinese Academy of Sciences (CAS) President's
International Fellowship Initiative (PIFI) (Grant No. 2018DM0034),
by the VolkswagenStiftung (Grant No. 93562), and by the EU Horizon 2020
research and innovation programme, STRONG-2020 project under grant
agreement No. 824093.

%

\end{document}